\documentclass[twocolumn,twoside,slac_two]{revtex4}
\usepackage{graphicx}
\usepackage{fancyhdr}
\newcommand{\rmmat}[1]{{\hbox{\rm #1}}}
\newcommand{\rmscr}[1]{{\rmmat{\scriptsize #1}}}

\begin{document}

\title{Magnetars}

%

\author{J. S. Heyl}
\affiliation{Department of Physics and Astronomy, University of
  British Columbia, 
Vancouver BC V6T 1Z1, Canada}

\begin{abstract}

Ultramagnetized neutron stars or magnetars are magnetically powered
neutron stars.   Their strong magnetic fields dominate the physical
processes in their crusts and their surroundings.   The past few years
have seen several advances in our theoretical and observational
understanding of these objects.   In spite of a surfeit of
observations, their spectra are still poorly understood.  I will
discuss the emission from strongly magnetized condensed matter
surfaces of neutron stars, recent advances in our expectations of the
surface composition of magnetars and a model for the non-thermal
emission from these objects.
\end{abstract}

\maketitle

\thispagestyle{fancy}


\section{INTRODUCTION}

Simply put magnetars are neutron stars whose magnetic fields dominate
their emission, evolution and manifestations.  In the late 1970s and
early 1980s,  a fleet of sensitive detectors of high-energy radiation
uncovered two new phenomena, the soft-gamma repeater and the anomalous
x-ray pulsar.  Strongly magnetized neutron stars provide the most
compelling model for both types of object, and observations over the
past few years indicate that these phenomena are two manifestations of
the same type of object.   Soft-gamma repeaters exhibit quiescent
emission similar to that of anomalous x-ray pulsars
\cite[e.g][]{1994Natur.368..432R,1994Natur.368..127M,1996ApJ...463L..13H,1999ApJ...510L.111H},
and anomalous x-ray pulsars sometimes burst
\cite{2002Natur.419..142G,2003ApJ...588L..93K}.
What makes magnetars a hot topic of
research is the rich variety of physical phenomena that strong
magnetic fields exhibit.

This article will focus on the quiescent emission from these
interesting objects rather than the bursts themselves (The reader may
wish to refer to the seminal work of Thompson and Duncan \cite{Thom95} for
details of the burst but may also want to look at Heyl and Hernquist
\cite{Heyl03sgr} for an alternative).  Furthermore, the article will
concentrate on recent results.

\section{MAGNETAR MANIFESTIONS}

\subsection{Soft-Gamma Repeaters}
\label{sec:soft-gamma-repeaters}

The first soft-gamma repeater was discovered on March 5, 1979 when
gamma ray detectors on nine spacecraft across our solar system
recorded an intense radiation spike \cite{1979Natur.282..587M}. 
The burst of gamma rays
originated from near a supernova remnant known as N49 in the Large
Magellanic Cloud.  The tail of the gamma-ray burst exhibited an
eight-second pulsation (in contrast with the classical gamma-ray
bursts which show no periodicities).   If one combined this
eight-second period with the age of the supernova remnant and assumed
that the object (presumably a neutron star) was born spinning much
faster, one estimated a magnetic field on the object of $\sim
10^{15}$~G, much larger than any neutron star discovered up to that point.

This first SGR became known as SGR 0526-66.   During 1979, 
two others were discovered: SGR 1806-20 (27 December 2004 event),
SGR 1900+14 (28 August 1998 event) \cite{1981Ap&SS..80....3M}.
In 1998 SGR 1627-41 became the
fourth SGR to be discovered \cite{1999ApJ...519L.139W}.

Thompson \& Duncan \cite{Thom95,Thom96} argued that the evolution of a
ultrastrong magnetic field could explain both the outbursts and
quiescent emission from soft-gamma repeaters, and the term ``magnetar'' was
born.  They argued earlier that if protoneutron star was born spinning
sufficiently rapidly, a dynamo could dramatically amplify the standard
pulsar magnetic field ($\sim 10^{12}$~G) that the protoneutron star
was born with to $\sim 10^{15}$~G or more \cite{Thom93b}.

\subsection{Anomalous X-ray Pulsars}
\label{sec:anomalous-x-ray}

1E~2259+586 was the first anomalous x-ray pulsar to be discovered
\cite{1981Natur.293..202F}.  In the early nineties these objects began
to form a unique class \cite{Mere95,Corb95,VanP95,Vasi97b}.   They
were found to have much in common with the soft-gamma repeaters.  

These objects typically have pulsed X-ray emission with steadily
increasing periods of several seconds, X-ray luminosities of
$\sim 10^{35}-10^{36}$~ergs~s$^{-1}$, soft spectra, and no detected
companions or accretion disks. Furthermore, they are
typically observed through hydrogen column densities
of $\sim 10^{22}$~cm$^{-2}$, indicating that they are not 
common. 

Two of the five confirmed AXPs are located near the centres of
supernova remnants 1E~2259+586 and
1E~1841-045\cite{1981Natur.293..202F,Vasi97b} as well as the AXP
candidate AX~J1845-0258 \cite{2000ApJ...542L..49V}.  The remaining
objects are 4U~0142+61 \cite{1994ApJ...433L..25I}, 1E~1048.1-5937
\cite{1986ApJ...305..814S}, 1RXS J170849.0-400910
\cite{1997PASJ...49L..25S} and possibly XTE J1810-197
\cite{2004ApJ...609L..21I}.

The emission from the AXPs fits neatly within the magnetar model
\cite{Thom96}.   Heyl \& Hernquist \cite{Heyl97kes} argued that the
thermal flux passing through an ultramagnetized, hydrogen or helium
envelope is sufficient to account for the x-ray emission from these
objects.  Hey \& Kulkarni \cite{Heyl98decay} examined how magnetic
field decay can augment the thermal energy budget.   For fields less
than about $10^{15}$~G magnetic field decay in a realistic model does
not strongly affect the emission from young (less than
10,000 years) AXPs but can greatly increase their lifetime as
observable x-ray sources.

Alternative models such as accretion have fallen by the wayside
because even very low mass companions have not been discovered
orbiting these neutron stars nor has the tell-tale optical emission
from even a truncated accretion disk been detected.

\subsection{Strongly Mangetized Radio Pulsars}
\label{sec:strongly-mang-radio}

Moreover, a number of radio pulsars have been discovered with inferred
magnetic field strengths similar to those of magnetars and apparently
exceeding the value $B_\rmscr{QED} \approx 4.4 \times 10^{13}$~G
\cite{2000ApJ...541..367C,2002MNRAS.335..275M,2003ApJ...591L.135M} It
is not clear why these objects have magnetic fields comparable to
the AXPs but do not exhibit AXP-like emission.  The inferred magnetic
field of AXP~1E2259+586 is actually smaller than PSR J1847-0130, an
otherwise ordinary radio pulsar.  The AXPs and SGRs exhibit much
greater x-ray emission than their relatively inactive cousins.  
\S~\ref{sec:phys-strong-magn} speculates further on the possible
differences between these two types of objects.

\section{PHYSICS IN A STRONG MAGNETIC FIELD}
\label{sec:phys-strong-magn}

Isolated neutron stars generally drawn on several sources of energy.
For most observed neutron stars, the dominant source of energy is the
rotation of the star.
\begin{equation}
E_\rmscr{rot} = \frac{1}{2} I \Omega^2 \approx 2 \times 10^{46} I_{45} \left (
 \frac{P}{1~\rmmat{sec}} \right )^{-2} ~\rmmat{erg}
\label{eq:1}
\end{equation}
Magnetars are typically young, highly magnetized neutron stars.   The
magnetic energy,
\begin{equation}
{\cal M} \, = \, {1\over{8\pi}} \, \int B^2 \, dV \, \approx 
2\times 10^{47} \, R_6^3\, B_\rmscr{NS,15}^2
 \,\,\,  {\rm ergs},
\label{eq:2}
\end{equation}
exceeds the rotational energy by an order of magnitude.  Nearly as
large is the thermal energy of the star
\begin{equation}
U \, \sim \, 10^{47} \, R_6^2 \, M_{1.4}^{1/3} \,
T_{8.5}^2 \,\,\,\, {\rm ergs},
\label{eq:3}
\end{equation}
where $M_{1.4}$ and $T_{8.5}$ are the mass 
and core temperature of the star in units of $1.4M_\odot$ and
$10^{8.5}$ K, respectively, and $R_6$ is the radius in units of 10~km
and $B_{\rmmat{NS},15}$ is the surface magnetic field of the star in
units of $10^{15}$~G.   Much more weakly magnetized objects such as
AXP~1E2259+586 act like magnetars as well.  For this object $U \sim
100{\cal M}$.  

It is unclear why some strongly magnetized neutron stars behave like
``magnetars'' rather than just happily spinning down and radiating as
a radio pulsar.  Not only do the soft-gamma repeaters and anomalous
x-ray pulsars share a penchant for bursting \cite[and Kaspi in these proceedings]{2003ApJ...588L..93K}
that the pulsars lack they also generally have much stronger x-ray
emission that radio pulsars -- they are not only strongly magnetized
but hot as well.  As we have seen, for AXP~1E2259+586 the thermal
energy exceeds a naive estimate of the magnetic energy by two orders
of magnitude.  It is quite natural to speculate that the heat flux
travelling through the surface layers of the AXPs and SGRs may play an
important role in their ``magnetar'' behaviour.  A thermomagnetic
interplay \cite{Blan83} between the strong magnetic field and the
strong heat flux may help to power their bursts .

Gamma rays from by far the largest burst from a soft-gamma repeater
arrived at Earth ten short days after the close of this meeting.  
The recent superbursts from SGR~1806-20 
\cite{2005astro.ph..3030P,2005astro.ph..2577M,2005astro.ph..2541M}
brings many of these issues to the fore.   Particularly the energy of
this burst has been estimated to be $\approx 2\times 10^{46}$~erg
\cite{2005astro.ph..3030P}, one hundred times that of the previous two
superbursts from soft-gamma repeaters.
   It has been argued that this
is a once in a century event \cite{2005astro.ph..3030P}.  Given that
detectors sensitive to such phenomena have only existed for 48 years
(if one generously assumes that Sputnik and the Explorer satellites
would have noticed such a burst), a once per century rate is rather
conservative.  To account for the number of observed SGRs, the
soft-gamma repeaters must have a lifetime of several thousand years,
so they would be expected to exhibit several dozens of such bursts.
The budget of magnetic energy alone is insufficient to account for
these bursts, so one is forced to conclude that either we are seeing
the very end of the SGR phase for SGR~1806-20 or that the energy
reservoir of SGR~1806-20 exceeds the magnetic energy stored in the
dipole field.

Subsequent observations of SGR~1806-20 have reduced the estimate for
its distance \cite{2005astro.ph..3171M}, but they still only obtain a
lower limit to the distance.   This result may reduce the energy budget
for the event and ease the energy crisis.  Further work is clearly
needed. 

Until the past year the observed quiescent emission from SGRs and AXPs
was thought to be dominated by thermal emission.  The following
subsections discuss the various physical processes that determine the
thermal emission from AXPs and SGRs (and possibly the non-thermal
emission as well).

\subsection{Vacuum Physics}
\label{sec:vacuum-physics}

For many purposes, vacuum polarization effects can be calculated using
an effective Lagrangian for the electromagnetic field.  Following the
usual convention, we write
\begin{equation}
{\cal L} \, = \, {\cal L}_0 \, + \, {\cal L}_1 \, + \, \cdots \, .
\label{eq:4}
\end{equation}
Here, ${\cal L}$ is the full Lagrangian density, ${\cal L}_0$ is
the classical term, and ${\cal L}_1$ includes vacuum corrections
to one-loop order.  Higher order radiative corrections would be
described by additional terms. The second-order term is smaller that
${\cal L}_1$ \citep{Ditt85}.

Because the Lagrangian is Lorentz invariant, both ${\cal L}_0$ 
and ${\cal L}_1$ can be expressed in terms of the Lorentz invariants
of the electromagnetic field
\begin{equation}
I \equiv F_{\mu \nu} F^{\mu \nu} = 2\left ( |{\bf B}|^2 \, - \,
|{\bf E}|^2 \right )
\label{eq:5}
\end{equation}
and
\begin{equation}
K\equiv \left [ \epsilon^{\lambda \rho \mu \nu} 
F_{\lambda \rho} F_{\mu \nu} \right ]^2 = 
-4\left ( {\bf E} \cdot {\bf B} \right )^2 ,
\label{eq:6}
\end{equation}
where $\epsilon^{\lambda \rho \mu \nu}$ is the completely 
antisymmetric Levi-Civita tensor.
The effective Lagrangian of
the electromagnetic field was derived by 
Heisenberg and Euler \cite{Heis36} and Weisskopf \cite{Weis36} using electron-hole
theory.  In rationalized Gaussian units,
we can write ${\cal L}_0$ and ${\cal L}_1$ as
\begin{equation}
{\cal L}_0 \, = \, - {{1}\over 4} \, I \, ,
\label{eq:7}
\end{equation}
\begin{eqnarray}
{\cal L}_1 \, &=& \, {{\alpha}\over{2\pi}} \, \int_0^{\infty} \,
{\rm e}^{-\chi} \, {{d\chi}\over{\chi ^3}} \, \Biggr [ \nonumber \\
& & ~~~  
i \, \chi^2 \, {{\sqrt{-K}}\over 4} \, {{\cos(J_+ \, \chi) +
\cos(J_- \, \chi)}\over {\cos(J_+ \, \chi) - \cos(J_- \, \chi)}}
\, + 
\\ 
& & ~~~ \, B_\rmscr{QED}^2 \, + \, I \, {{\chi^2}\over 6} \, \Biggr ]
,
\nonumber
\end{eqnarray}
where
\begin{equation}
J_{\pm} \, \equiv \, {1\over{2 B_\rmscr{QED}}} \, \left [
-I \, \pm \, i \, \sqrt{-K} \right ]^{1/2} ,
\label{eq:8}
\end{equation}
$\alpha \equiv e^2/\hbar c$ is the fine structure constant,
$B_\rmscr{QED} \equiv m^2c^3/e\hbar \approx 4.4\times 10^{13}$ G, and
a similar quantity can be defined for the electric field,
$E_\rmscr{QED} \equiv m^2c^3/e\hbar \approx 2.2\times 10^{15}$ V/cm.
The above expressions for ${\cal L}_0$ and ${\cal L}_1$ are identical
to the corresponding terms in eq. (45a) of Heisenberg \& Euler \cite{Heis36}.

The above integral cannot be evaluated explicitly, in general.  Heyl
\& Hernquist \cite{Heyl97hesplit} have derived an analytic
expression for ${\cal L}_1$ as a power series in $K$:
\begin{equation}
{\cal L}_1 \, = \, {\cal L}_1 (I,0) \, + \, K \, {{\partial {\cal L}_1}
\over{\partial K}} \Bigg |_{K=0} \, + \, \cdots \, ,
\label{eqnL}
\label{eq:9}
\end{equation}
where the first term in this series is
\begin{eqnarray}
{\cal L}_1 (I,0)  &=&  {\alpha\over{4\pi}} \, I \, X_0 \left ( 
{1\over \xi} \right )  \\
&=& {\alpha\over{4\pi}} 
\int_0^\infty  {\rm e}^{-u/\xi} {{du}\over{u^3}} 
\left ( -u \coth u  +  1  +  {{u^2}\over 3} \right ) ,
\nonumber \\
& & 
\end{eqnarray}
and
\begin{equation}
\xi \, = \, {1\over B_\rmscr{QED}} \, \sqrt{{I\over2}} =
\frac{B}{B_\rmscr{QED}},\,
\label{eq:10}
\end{equation}
where the final equality holds if the electric field can be neglected.

The function $X_0$ can be evaluated analytically (as can the
higher order terms in the expansion for ${\cal L}_1$) with
the result \cite{Heyl97hesplit}
\begin{eqnarray}
X_0(x) \, &=& \, 4\int_0^{x/2-1} \, \ln (\Gamma(v+1)) \, dv \, - \,
{1\over 3} \, \ln x \, + \, {\cal C} \, 
\nonumber \\
& & ~~~ - \, \left [ 
1 \, + \, \ln \left ( {{4\pi}\over x} \right ) 
\right ] \, x \, 
 \\
& & ~~~
+ \, \left [ {3\over 4} \, 
+ \, {1\over 2} \, \ln \left ( {2 \over x} \right ) \right ] \, x^2 \,
,
\nonumber 
\end{eqnarray}
where
\begin{equation}
{\cal C} \, = \, 2\ln 4\pi \, - \, 4\ln A \, - {5\over 3} \ln 2 \, = \, 
2.911785285,
\label{eq:11}
\end{equation}
and
the constant $\ln A$ is related to the first derivative of the
Riemann zeta function, $\zeta^{(1)}(x)$, by
\begin{equation}
\ln A \, = \, {1\over {12}} \, - \, \zeta^{(1)} (-1) \, = \,
0.248754477 \, .
\label{eq:12}
\end{equation}
The integral of $\ln \Gamma (x)$ can be expressed in terms of
special functions (eqs. 18, 19 in \cite{Heyl97hesplit}).  (See also Dittrich 
et al. 1979; Ivanov 1992, but note the cautionary remark
in \cite{Heyl97hesplit}.)

The expression above for $X_0(x)$ can be expanded in either
a Taylor series in the weak field limit, $\xi \ll 1$, or an
asymptotic series in the strong field limit $\xi \gg 1$, as
can the higher order terms in equation (\ref{eqnL}), to give
series expansions for ${\cal L}_1$ as a function of either $I$ and
$K$, or equivalently $B$ and $E$.  In particular, to lowest
order in the weak field limit ($\xi \ll 1$)
\begin{equation}
{\cal L}_1 \, = \, {{\alpha}\over{90 \pi}} \, {1\over {B_\rmscr{QED}^2}}
\left [ (B^2 \, - \, E^2)^2 \, + \, 7 ({\bf E} \cdot {\bf B})^2 
\right ] \, + \, \cdots 
\label{HEL1}
\label{eq:13}
\end{equation}
In the limit of an ultrastrong magnetic field, $B \gg B_\rmscr{QED}$,
but for a weak electric field, $E \ll E_\rmscr{QED}$, 
${\cal L}_1$ can be written
\begin{equation}
{\cal L}_1 \, = \, {{\alpha}\over{6 \pi}} \, B^2 \,
\left [ \ln \left ( {B\over {B_\rmscr{QED}}} \right ) \, - \, 12 \ln A
\, + \, \ln 2 \right ] \, + \, \cdots \, 
\label{HEAL1}
\label{eq:14}
\end{equation}
(see e.g. eq. 29 in \cite{Heyl97hesplit} for the higher order terms).  We note that
equations (\ref{HEL1}) and (\ref{HEAL1}) agree, respectively,
with the corresponding terms in eqs. (43) and (44) of
\cite{Heis36}.  The analysis of \cite{Heyl97hesplit} generalizes the
expressions of Heisenberg \& Euler to arbitrary order.

Also, note that while our expression for ${\cal L}_0$ is identical 
to eq. (4-120) of Itzykson
\& Zuber \cite{Itzy80}, equation (\ref{HEL1}) 
differs from their eq. (4-125) by a factor of 
$1/4\pi$, as a consequence of
a difference in the system of units employed.\footnote{Itzykson
\& Zuber \cite{Itzy80} use Heaviside's units in defining the Coulomb
force; $E^2$ and $B^2$ are smaller in this system than in ours by
a factor of $4\pi$.}  Our expressions for 
${\cal L}_0$ and ${\cal L}_1$ both differ from those in
Berestetskii et al. \cite{Bere82} by an overall factor of $1/4\pi$ 
(their eqs. 129.2 and 129.21); however, the dynamics of the fields
are invariant with respect to rescalings of the Lagrangian.

We emphasize that the expression for the Lagrangian in the weak-field
limit, equation (\ref{HEL1}), cannot be applied to magnetar fields
which are thought to have $B_\rmscr{NS} \gg B_\rmscr{QED}$.  The use of the
weak-field expressions to calculate, e.g. the index of refraction of
the vacuum near the surface of a magnetar will result in estimates
that are incorrect by more than an order of magnitude at the relevant
field strengths.  In this limit, the Lagrangian should instead be
approximated by e.g. equation (\ref{HEAL1}), which is an asymptotic
series for ${\cal L}_1$ valid for $B \gg B_\rmscr{QED}$ and $E \ll
E_\rmscr{QED}$.

After integrating out the effects of the virtual electron-positron
pairs, one is left with an effective Lagrangian and many
electromagnetic phenomena can be treated by simply treating the vacuum
itself as a medium endowed whose properties are described by the
Lagrangian.   This approximation is valid for photon energies much less
than $m_e c^2$.  For higher energy photons the assumption of a
homogeneous field implicit in the above construction of the effective
Lagrangian breaks down and one must resort to other methods \cite[for example][]{Schw51}.

\subsection{Atomic Physics}
\label{sec:atomic-physics}

The atomic physics of materials on the surfaces of magnetars is
dominated by the magnetic field rather that the atomic nucleus.   A
useful figure of merit is the ratio of the cyclotron energy of an
electron to the binding electron of an electron to a nucleus of atomic
number $Z$
\begin{eqnarray}
\frac{E_\rmscr{cyclo}}{E_\rmscr{Rydberg}} 
&=& \frac{m_e c^2 B/B_\rmscr{QED}}{Z^2 / 2 \alpha^2 m_e c^2} = \frac{2
  B}{\alpha^2 Z^2 B_\rmscr{QED}} \\
&=& \frac{B}{1.1 Z^2 \times 10^{9}~\rmmat{G}}
\end{eqnarray}
where $B_\rmscr{QED}$ is defined in \S\ref{sec:vacuum-physics}.  We
see for magnetars this ratio is large for $Z$ up to one hundred or
more.  The atomic physics is in this regime has been probed
extensively for hydrogen and helium \cite[for
example][]{Rude94,Heyl98atom} but approximately for larger atoms such
as iron \cite{1991MNRAS.253..107M}.

What remains uncertain is how these atoms interact and what are the
properties of strongly magnetized material in the bulk at low
pressure.  Is it a solid, liquid or gas?  Is it metallic?  These
questions are crucial to understand the details of the thermal
emission from neutron stars.  For further details on the physics of
atoms and molecules in strong magnetic fields the reader is urged to
consult the excellent review of Lai \cite{2001RvMP...73..629L}.

\subsection{Thermal Physics}
\label{sec:thermal-physics}

Through much of the envelope of a magnetar the electron gas is
effectively one-dimensional because only the first Landau level is
typically filled \cite{Heyl97analns,Heyl98numens}.  This restriction 
dramatically affects the thermodynamic properties of the material and
allows approximate yet reasonably accurate analytic treatment of the
thermal conduction through the envelope of a magnetar
\cite{Heyl97analns}.

The key parameters are the chemical potential of the electrons, the
temperature and the strength of the magnetic field.   It is helpful to
define the dimensionless quantities,
\begin{equation}
 \zeta =
\frac{\mu_e}{m_e c^2},\, x^2=\zeta^2-1,\, \tau=\frac{k T}{m_e c^2}
\label{eq:15}
\end{equation}
in addition to $\xi=B/B_\rmscr{QED}$ from \S~\ref{sec:vacuum-physics}.
The number density of electrons in the degenerate limit ($\tau \ll\
\zeta - 1$) is given by
\newcommand{\lame}{{\lambda\!\!\!\!\hspace{0.2mm}^-}}
\begin{equation}
n_{e,0} = \frac{1}{3\pi^2\lame^3} x^3
\label{eq:16}
\end{equation}
for a weakly magnetized gas ($x \gg\ 2\xi$) and
\begin{equation}
n_{e,\rmscr{magnetar}} = \frac{\xi}{2\pi^2\lame^3}  x
\label{eq:17}
\end{equation}
for a strongly magnetized gas.  The symbol $\lame$ denotes the Compton
wavelength of the electron, $\hbar/(m_e c)$.

The pressure of the magnetized gas also differs from the unmagnetized
case.
The number density of electrons in the degenerate limit ($\tau \ll\
\zeta - 1$) is given by
\begin{equation}
P_{e,0} = \frac{1}{8\pi^2} \frac{m_e c^2}{\lame^3}
\left [ x \zeta \left( \frac{2}{3} \zeta^2 - \frac{5}{3} \right ) +
  \ln ( x + \zeta ) \right ]
\label{eq:18}
\end{equation}
for a weakly magnetized gas ($x \gg\ 2\xi$) and
\begin{equation}
P_{e,\rmscr{magnetar}} = \frac{\xi}{4\pi^2} \frac{m_e c^2}{\lame^3} \left [ x \zeta - \ln ( x+\zeta )
  \right ]
\label{eq:19}
\end{equation}
for a strongly magnetized gas. Although the magnetic field changes the
pressure dramatically, the electric pressure remains isotropic due to
the changing mangetization as the electrons are compressed
\cite{Hern85}.

An important comparison is at what density do the electron become
degenerate in the two cases.   Magnetar envelopes are generally cool,
$\tau \ll 1$ so $\zeta - 1 \approx x^2/2 \approx \tau$ at the onset of
degeneracy.   The ratio of the densities of the gas at the onset of
degeneracy in unmagnetized and the ultramagnetized case is given by
\begin{equation}
\frac{n_{e,\rmscr{magnetar}}}{n_{e,0}} =  \frac{3}{2}
\frac{\xi}{x^2} = \frac{3}{4} \frac{\xi}{\tau} \gg 1.
\label{eq:20}
\end{equation}
The ratio of the pressures at the onset of degeneracy is given by a
similar factor.  For a magnetar, $\xi \sim 10 - 100$ and $\tau \sim
10^{-1}$, yielding a factor of 100$-$1000 increase in the density,
column density and pressure at the onset of degeneracy.  Magnetar
envelopes only become degenerate densities larger than
\begin{eqnarray}
\rho_\rmscr{ND/D} &=& 3.92 \times 10^5 \rmmat{ g cm}^{-3} \xi 
( \ln b_\rmscr{Typical} )^{1/7} 
\nonumber \\
& & ~~~ \times 
A_{56}^{6/7} Z_{26}^{-5/7} T_\rmscr{eff,6}^{4/7}
g_{s,14}^{-1/7}.
\label{eq:21}
\end{eqnarray}
where $b_\rmscr{Typical}$ is a typical value of $\xi/\tau$ in the
envelope, $b_\rmscr{Typical} \approx 6 \times 10^3 \xi$.  The
unmagnetized envelope become degenerate at densities $\sim 10^2 -
10^4$~g~cm$^{-3}$ \cite{Hern84b}.

\section{THERMAL EMISSION FROM MAGNETARS}
\label{sec:therm-emiss-from}

The thermal emission from magnetars is powered by a combination of the
residual heat from the supernova explosion (Eq.~\ref{eq:3})and
possibly the magnetic field (Eq.~\ref{eq:2}).  The accretion
alternative predicts more optical emission than is observed from
these objects \cite[but see][for an alternative point of
view]{2003ApJ...599..450E}.  If the thermal energy originates from
within the star, the energy must first travel through the insulating
envelope of the neutron star.  The conductivity of this layer
throttles the heat flow and determines the total thermal emission from
these objects.

In the non-degenerate regime, the conductivity is nearly constant and
determined by the outgoing heat flux both for magnetars and weakly
magnetized neutron stars \cite{Hern84b,Heyl97analns}.  The photons in
the extraordinary polarization mode dominate the heat conduction, and
the conductivity itself is strongly dependent on the magnetic field.
In the degenerate regime the conductivity increases dramatically.
Along the field lines, the strong magnetic field strengthens the heat
conduction, while across the field lines, the field inhibits it.  For
a fixed internal temperature of the star, the emerging flux is
approximately proportional to $Z^{-0.7} B^{0.4} \cos^2 \psi$ where
$\psi$ is the angle between the local magnetic field ($B$) and the
normal and $Z$ is the atomic number of the nuclei that comprise the
envelope \cite{Heyl97analns,Heyl97kes}.  Potekhin et
al. \cite{2003ApJ...594..404P} examined the dependence of the cooling
of neutron stars (including the superfluidity of their interiors) on
the presence of a light-element envelope in further detail, verifying
the earlier conclusions.

\subsection{Magnetized Atmospheres}
\label{sec:magn-atmosph}

The study of mangetized neutron stars is simplified by the fact that
neutron stars atmospheres are exceedingly thin.  Unfortunately, this
is an inadequate consolation for the complications that the strong
magnetic fields introduce.
\begin{itemize}
\item  
As discussed in \S~\ref{sec:atomic-physics} the binding energies of
even light species such as hydrogen and helium may exceed several
hundred electron volts in the magnetar fields, so magnetar atmospheres
are probably only partially ionized \cite[see Figs.~1 and~2
of][]{2004ApJ...600..317P}.  The light element atmospheres of more
weakly magnetized stars are typically fully ionized, but even in weak
fields heavier species such as carbon, oxygen and others will not be
fully ionized.  The presence of bound species affects not only the
opacity of the plasma but also its equation of state and
electromagnetic polarization.
\item
The opacities for the two polarization modes differs by several orders
of magnitude, so the polarized radiative transfer must be calculated.
\item 
Vacuum polarization (\S~\ref{sec:vacuum-physics}) complicates the
behaviour of the propagation modes of the plasma itself.  As radiation
propagates through the vacuum resonance where the vacuum and plasma
contribute equally to the index of refraction, the mode that below the
resonance had the higher opacity now has the lower opacity and vice
versa.  This effect complicates the radiative transfer for for
radiation slightly above the peak of the thermal distribution
\cite{2002ApJ...566..373L,2003ApJ...583..402O}.
\item
The opacites depend strongly on the direction of that the radiation
propagates relative to the magnetic field.   Where the field is
normal to the surface only the altitude is important (like in
unmagnetized transfer) but the distribution must be sampled more
finely to handle the sensitive angular dependence.   Where the field
is makes an angle with respect to the normal (the general case), the
intensity must be calculated both in altitude and azimuth.
\end{itemize}

\begin{figure}
\includegraphics[width=80mm]{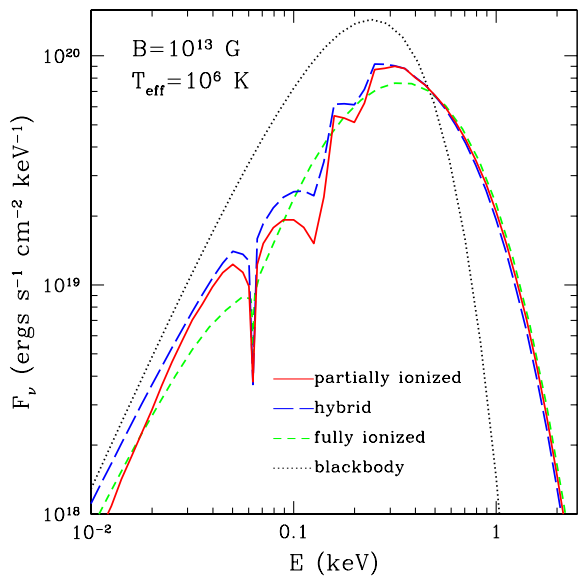}
\caption{The total emergent spectrum from a strongly magnetized
  neutron star atmosphere from the calculations of Potekhin et al.
  \cite{2004ApJ...612.1034P}.  A blackbody
  spectrum is compared with several atmospheric calculations of
  increasing realism.  The fully ionized model assumes that neutral
  species are absent.  The hybrid model includes 
  partial ionization to calculate the equation of state and the opacity
  but assumes that polarization modes are those of a fully ionized
  plasma.   The partially ionized model includes the effects of
  partial ionization on the opacity, equation of state and
  polarization.  All of the models include vacuum polarization}
\label{fig:pote}
\end{figure}

The recent work of Potekhin et al.\cite{2004ApJ...612.1034P} may be
considered an example of the state of the art in magnetized
atmospheres.  Fig.~\ref{fig:pote} depicts the results for calculations
that fully include the effects of partial ionization and vacuum
polarization as well as spectrum using less realistic assumptions.
The shape of the spectrum changes dramatically as the calculations
become more sophisticated.   Even from these comprehensive
simulations, several important effects are absent.  First, the
magnetic field is assumed to pointing in the normal direction.
Relaxing this constraint makes the calculations more onerous but
similar in principle and is required to simulate the emission from the
entire surface of the neutron star.  Second, the magnetic field is
still two orders of magnitude less than that of magnetars.   As the
magnetic field increases the difference in the opacities between the
two modes increases and also the material in the outer layers may
undergo a phase transition \cite{2004astro.ph..6001V}.   Finally the
surface layers may not consist of hydrogen but rather helium or
heavier elements (see \S~\ref{sec:diffusion}).

A tantalizing alternative is that magnetar atmospheres aren't
atmospheres at all, but rather the surface of a magnetar is condensed.
The condensation temperature may be as high as $10^6$~K for iron in
fields stronger than $10^{13}$~G and even for hydrogen in fields of a
few $\times 10^{14}$~G.  In this situation detailed balances allows
the emergent spectrum from the star to be determined by understanding
how light reflects off of it.  Van Adelsberg et
al. \cite{2004astro.ph..6001V} calculate emergent thermal spectra from
the condensed surface of a magnetar.   If the surface is smooth, the
spectrum deviates from a blackbody by at most a factor of two and
exhibits only mild absorption features that correspond to the ion
cyclotron resosonace and the plasma frequency within the surface.
The emission for rough surfaces resembles that of a blackbody more
closely. 
\begin{figure}
\includegraphics[width=80mm]{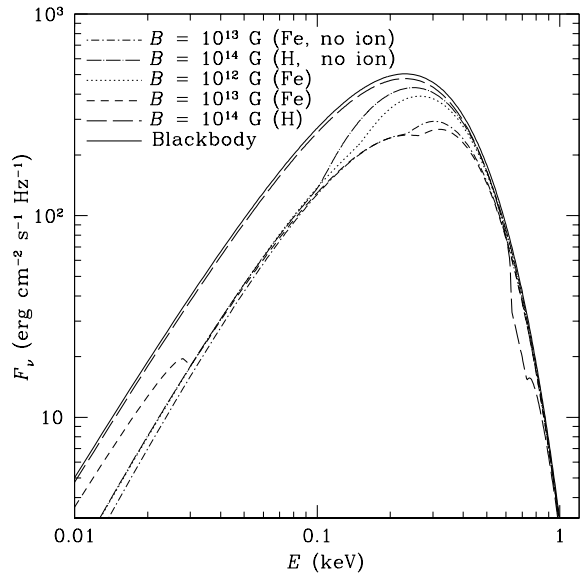}
\caption{The total emergent spectrum from a strongly magnetized
  condensed neutron star atmosphere from the calculations of 
  van Adelsberg et al. \cite{2004astro.ph..6001V}.}
\label{fig:vanA}
\end{figure}

\subsection{Magnetospheric Propagation}
\label{sec:magn-prop}

Regardless of the detailed composition of the atmosphere of magnetars,
the vast difference between the opacities in the two polarization
modes nearly ensures that the radiation will emerge from the
atmosphere highly polarized perpendicular to the direction of the
local magnetic field.  Pavlov and Zavlin \cite{2000ApJ...529.1011P}
argued that because the direction of the magnetic field varies across
the surface of the neutron stars, the net polarization from the entire
visible surface is greatly diminished.  This argument neglects that
the region surrounding the magnetar is optically active.  Specifically
at the x-ray energies, the the intense magnetic field decouples 
the propagation modes in the magnetosphere through vacuum polarization
(\S~\ref{sec:vacuum-physics}).   Heyl and Shaviv \cite{Heyl01qed}
argued that polarized radiation propagates adiabatically through the
magnetosphere.  The adiabatic condition holds within the
polarization-limiting radius  \cite[see][for the plasma analogue]{Chen79},
\begin{eqnarray}
 r_\rmscr{pl} &\equiv& \left ( \frac{\alpha}{45}
 \frac{\nu}{c} \right )^{1/5} \left ( \frac{\mu}{B_\rmscr{QED}} \sin
 \beta \right )^{2/5} \\ \nonumber &\approx & 1.2 \times 10^{7} \left
 ( \frac{\mu \sin \beta}{10^{30}~\rmmat{G cm}^3} \right )^{2/5}
\left (
 \frac{\nu}{10^{17}~\rmmat{Hz}} \right)^{1/5} \rmmat{cm},
\end{eqnarray}
where $r$ is the distance from the center of the star, $\mu$ is the
magnetic dipole moment of the neutron star, and $\beta$ is the angle
between the dipole axis and the line of sight.  The observed
polarization reflects not the direction of the magnetic field at the
surface of the star but at the polarization-limiting radius;
consequently, if this radius is much larger than the radius of the
star, the polarized radiation from disparate regions of the stellar
surfaces adds conherently an a large net polarization results.

\begin{figure}
\includegraphics[width=80mm]{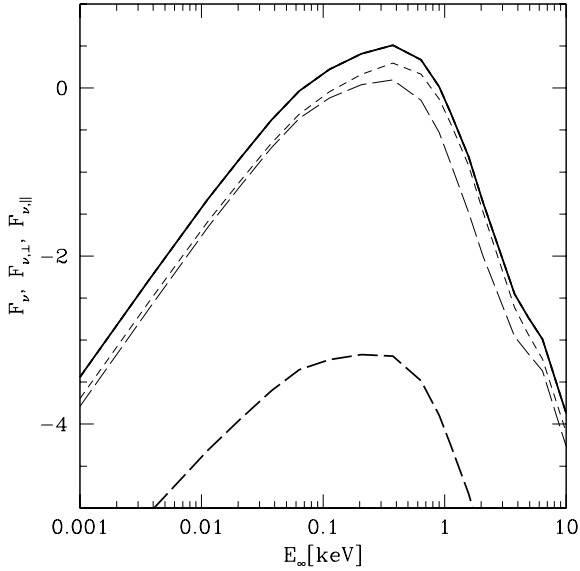}
\caption{The total emergent flux from the visible surface of the star.
The magnetic pole makes an angle of 60$^\circ$ with the line of sight.
The radius of the star is 12~km and its mass is 1.4~M$_\odot$.  The
effective temperature at the magnetic pole is $10^{6.5}$~K.  The
calculation is for $\mu=10^{32}$~G~cm$^3$, corresponding to a surface
field of $\sim 10^{14}$~G respectively.  The solid curve traces the
total flux.  The short dashed curve traces the flux polarized
perpendicular to the projection of the magnetic moment in the plane of
the sky.  The long-dashed curve traces the flux polarized parallel to
the projected magnetic moment.  The heavy curves trace the results
including vacuum polarization and the light curves neglect it.}
\label{fig:spectrumplot}
\end{figure}

Fig.~\ref{fig:spectrumplot} depicts the polarized spectrum summed over
the entire surface of the neutron star.  The local emergent radiation
spectrum is determined from an atmospheric model that assumes that the
plasma is fully ionized hydrogen and neglects vacuum polarization
within the atmosphere.  The bold curves trace the results from a
calculation that includes vacuum polarization within the magnetosphere
and the light curves neglect vacuum polarization.  If vacuum
polarization within the magnetosphere is included the radiation is
more than 99\% polarized near the peak of the spectrum.  Without it
the total emission is more modestly polarized by about 10\%.

In the X-ray regime, the radiation from the entire surface of the
neutron star is highly polarized for surface magnetic fields greater
than $10^{12}$~G or so.   At lower photon energies, especially in the
optical and infrared, the observed polarization is a sensitive probe
of the magnetic field and plasma density within the magnetosphere as
well as the radius of the neutron star \cite{Shan04}.   Polarimetry of
magnetars and neutron stars in general will provide a unique new probe
of their environments and interiors.

\subsection{Magnetic Field Decay}
\label{sec:magnetic-field-decay}

Thompson and Duncan \cite{Thom96} argued that the decay of the strong
magnetic field of a magnetar could account for the quiescent emission
from the surface and examine the gradual evolution of the magnetic
field of the star.  Goldreich and Reisenegger \cite{Gold92} examine
several modes of magnetic field decay: Ohmic diffusion, Hall drift and
ambipolar diffusion.  These processes have the following timescales:
\def\rhorhon{\left ( \frac{\rho}{\rho_\rmscr{nuc}} \right )}
\def\yr{\rmmat{~yr}}
\begin{eqnarray}
t_\rmscr{Ohmic} & \sim & 2 \times 10^{11} \frac{L_5^2}{T_8^2}
\rhorhon^3 \yr \\
t^\rmscr{s}_\rmscr{ambip} & \sim & 3 \times 10^{9} \frac{L_5^2
T_8^2}{B_{12}^2} \yr \\
t^\rmscr{irr}_\rmscr{ambip} & \sim & \frac{5 \times 10^{15}}{T_8^6
B_{12}^2} \yr + t^\rmscr{s}_\rmscr{ambip} \\
t_\rmscr{Hall} & \sim & 5 \times 10^{8} \frac{L_5^2
T_8^2}{B_{12}} \rhorhon \yr 
\end{eqnarray}
where $L_5$ is a characteristic length scale of the flux loops through
the outer core in units of $10^5$~cm, $T_8$ is the core temperature in
units of $10^8$~K and $B_{12}$ is the field strength in units of
$10^{12}$~G.  Ohmic decay dominates in weakly magnetized neutron stars
($B\lesssim 10^{11}$~G), fields of intermediate strength decay ($B
\sim 10^{12} - 10^{13}$~G) via Hall drift, and intense fields ($B
\gtrsim 10^{14}$~G) are mostly strongly affected by ambipolar
diffusion.  Thompson and Duncan \cite{Thom96} examined the possibility
of obtaining an equilibrium between neutrino cooling and heating
through magnetic field decay for $B \sim 10^{15}$~G and $T \sim
10^8$~K.

In the case of a magnetar older than a few hundred years, ambipolar
diffusion dominates the other porcesses at least in the core of the
neutron star.  Ambipolar diffusion involves motion of the charged
species (electrons, protons and possibly muons) relative the the
neutrons.  In the motion of the change species results a change in the
density of these particles (if the flow is irrotational), the chemical
equilibrium must be reestablished through weak interactions.
Otherwise the flow is solenoidal and weak reactions are not required.

\begin{figure*}
\includegraphics[width=80mm]{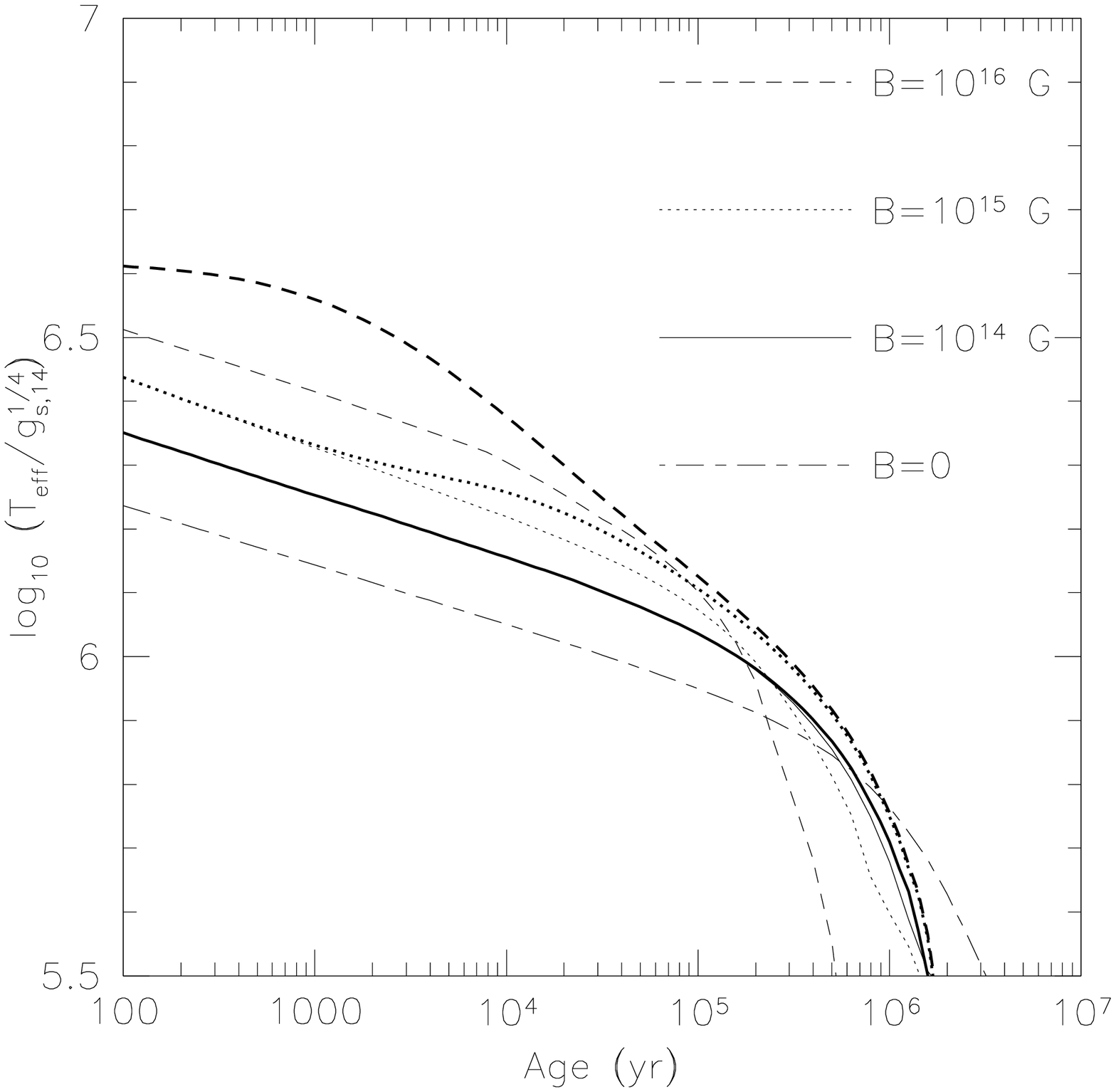}
\includegraphics[width=80mm]{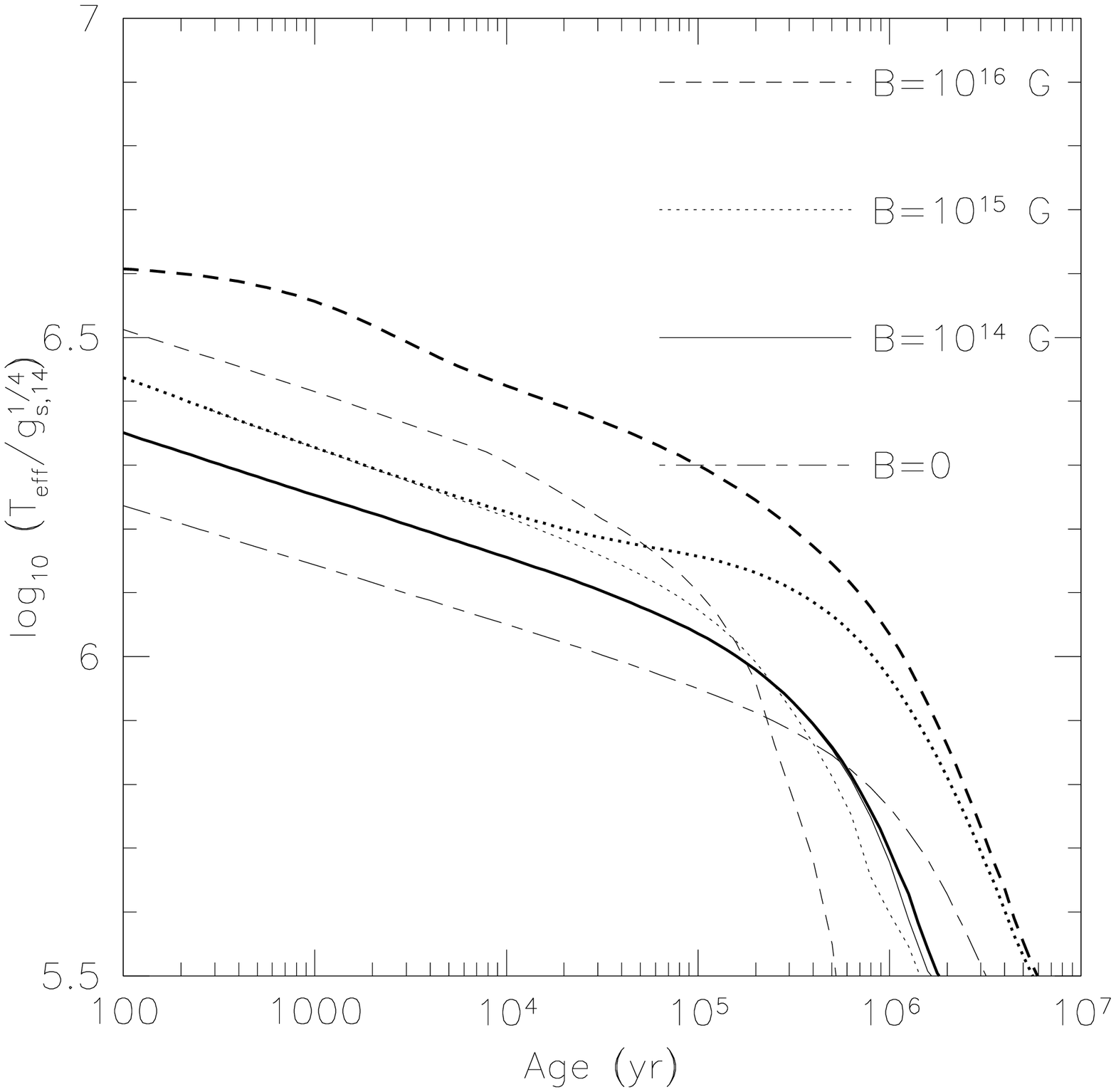}
\caption{The left panel depicts the evolution of the effective
temperature (left) and magnetic field (right)
magnetic field decays through the solenoidal mode.
The right panel depicts the results for decay through the irrotational
mode.
The bold solid, dotted and dashed lines give the results for
$B_i=10^{14}, 10^{15}$ and $10^{16}$~G respectively.  The light lines
give the results without any field decay, and the long-short-dashed
line gives the cooling evolution for an unmagnetized neutron star.}
\label{fig:decay}
\end{figure*}

Using the results of Goldreich and  Reisenegger
\cite{Gold92}, Heyl and Kulkarni \cite{Heyl98decay} examined how
magnetic field decay would affect the thermal emission from the
surface of a magnetar and its cooling, including a realistic treatment
of the strongly magnetized envelope of the neutron star.
Fig.~\ref{fig:decay} shows the results of their analysis.  If the
magnetic field is sufficiently strong ($B\sim 10^{16}$~G) field decay
has a dramatic effect on the emission from young magnetars, less than
ten thousand years old.  Even
without field decay the magnetic field dramatically increases the heat
flux through the surface of the neutron star.   For intermediate fields
($B\sim 10^{15}$~G) field decay does not strongly affect the emission
of the young magnetars, but for fields similar to and stronger than
$10^{15}$~G field decay increases the lifetime of hot magnetars by a
factor of several.   This dramatically increases the number of hot
magnetars that you would expect to see in the Galaxy and possibly the
evolution of soft-gamma repeaters and anomalous x-ray pulsars.

\subsection{Diffusion}
\label{sec:diffusion}

A crucial input for our understadning of the emission from thes
surface of neutron stars is the composition of their outer layers.
As discussed in \S~\ref{sec:thermal-physics} and
\S~\ref{sec:magn-atmosph} the presence of light elements in these
layers can have a dramaticially effect on the total thermal emissiokn
from the star as well as its spectrum.  The strong gravitiational
acceleration at the surface of the neutron star ($g \sim
10^{14}$~cm~s$^{-2}$) ensures that the lighter elements will float to
the top of the atmosphere; consequently, one would expect the emergent
spectrum and flux to be determined by the lightest elements present in
the outermost fluid layers of the star.   If any material fell back
onto the neutron star during or after the supernova, it spallates as
it hits the surface, resulting in hydrogen and helium nuclei.   

The thickness of a pure hydrogen layer is limited the stabilty of
 protons against decaying into neutrons.  If the chemical potential of
the electrons exceeds mass difference between a proton and a neutron
($\zeta \approx 2.53, x \approx 2.33$)
it becomes energetically favourable for protons to combine with
electrons and form neutrons.  These neutrons will quickly bind to
protons, forming deuterons and ultimately helium nuclei.

According to equations~\ref{eq:16} and~\ref{eq:17}, this occurs at a
density of about $1.2 \times 10^7 \rmmat{g cm}^{-3}$ for a weakly
magnetized gas and about $3.4 \xi \times 10^6 \rmmat{g cm}^{-3}$ for a
strongly magnetized gas in which $\xi \gg 1$.

The thickness of the hydrogen layer that can be supported on a
magnetar increases with the strength of the magnetic field.  The
thickness of a helium layer is assumed to be limited by the shell
flash instability that results in Type-I X-ray bursts on neutron
stars.  This criterion yields a maximum column density of a helium
layer of approximately $10^{8-9}$~g~cm$^{-2}$
\cite[e.g.][]{Nara03typei}.  It is unclear whether the strong magnetic
field affects nuclear burning of helium, but it does affect hydrogen
burning \cite{Heyl96fus}.  If a neutron star accretes more than about
$10^{22}$~g of material after its birth, one would expect find moving
from the outside inward hydrogen, helium and then carbon or other
heavier elements.

These results neglect the fact that although the layers are
stabily stratified, a thermally excited tail of protons will
penetrate the layer of helium and possibly be captured onto the carbon
and other nuclei lying below.  Chang, Arras and Bildsten 
\cite{2004astro.ph.10403C} examined this problem in detail.
As protons diffuse through the helium layer, their density drops as a
power law in the non-degenerate regime and exponetially in the
degenerate regime.   In a magnetar the electrons become degenerate at
a much higher column density than in a neutron star with a smaller
magnetic field (Eq.~\ref{eq:21}).   Although diffusive nuclear burning
is not terribly important over the timescale of a few thousand years
for weakly magnetized neutron stars, in magnetars a thin atmospheric
hydrogen layer is consumed as protons diffuse through a maximally thick
helium layer in only a few years.

The efficiency of diffusive nuclear burning on a magnetar indicates
that the atmospheres and enevlopes of magnetars likely consist of
helium or heavier elements rather than hydrogen.   Because ionized
hydrogen atmospheres has generally be the focus of research into
magnetar atmosphere (see \S~\ref{sec:magn-atmosph}) clearly more work
is needed.

\section{NON-THERMAL EMISSION FROM MAGNETARS}
\label{sec:non-thermal-emission}

Until the past year the quiescent emission from magnetars was thought
to be mostly due to thermal emission from the surface of the star.
Their soft x-ray spectra are notoriously well described by blackbody
models (possibly with a power-law component at high energies)
\cite[for example][]{Pern00axp}.   However, recent results from
INTEGRAL indicate that the thermal emission is just the tip of the
iceberg \cite[and Hurley in these
  proceedings]{2004astro.ph.11695M,2004ATel..293....1D,2004astro.ph.11696M}.  
The optical emission from AXPs is also clearly non-thermal (or at
least not from the surface) \cite{2004astro.ph..4144O}.   

Thompson and Beloborodov \cite{2004astro.ph..8538T} present two models
to account for the non-thermal hard x-ray emission from the AXP
1E~1841-045.  The first involves the heating of a surface layer of the
neutron star to $k T \sim 100$~keV by currents driven in the
magnetosphere by the twisting of the magnetic field in the crust.  The
second model focusses on a region far from the star (at ten stellar
radii) where the electron cyclotron resonance is approximately 1~keV.
The thermal flux from the surface exerts a force on the
current-carrying electrons in this region and a large electric field
develops.  A positron injected into this region quickly accelerates to
energies where it can upscatter keV photons above the threshold for
pair creation.  The pairs emit synchrotron radiation consistent with
the observed spectrum.

Heyl and Hernquist \cite{Heyl05sgr} present an alternative picture
that can account for both the optical and hard x-ray non-thermal
emission.  \S~\ref{sec:vacuum-physics} outlines the dynamics of the
electromagnetic field including radiative corrections.  A key result
is that the effective Lagrangian of QED makes electrodynamics
non-linear.   Specifically simple electromagnetic waves travelling
through a strong magnetic field develop regions when the electric and
magnetic field are discontinuous -- shocks \cite{Heyl98shocks}.
These shocks also form as fast waves travel through an
ultramagnetized, ultrarelativistic plasma \cite{Heyl98mhd}.   

As a fast wave propagates away from the surface of the neutron star
and forms a shock, the energy release in the shock powers the
formation of electron-positron pairs at rest in the frame of the
shock.  These pairs emit synchrotron photons that produce more pairs.
This pair cascade stops because the total optical depth for pair
creation strongly decreasing with the energy of the photons.  The
final generation of pairs has a typical Lorentz factor of $\gamma =
0.05 \xi^{-1}$ with $\xi \ll 1$ far from the surface of the star.
These pairs emit synchrotron photons with an energy ranging from
\begin{equation}
E_\rmscr{break}  = \gamma^2\hbar  \omega_B \approx 
2.5 \times 10^{-3} \xi^{-1} m c^2 .
\label{eq:23}
\end{equation}
down to $E_0 = \xi m c^2$.  At each radius the pairs cool in a fixed
magnetic field, yielding a spectrum $d E \propto E_\gamma^{-1/2} d
E_\gamma$ between the values of $E_0$ and $E_\rmscr{break}$ at the
innermost edge of the pair production region.  At energies greater
than $E_\rmscr{break}$ or less than $E_0$ the emission comes from
pairs produced further from the star.  Simulations of the fast-mode
breakdown in a dipole geometry \cite{Heyl05sgr} indicate that the fast
mode delivers equal amounts of energy in equal ranges of mangetic
field, yielding the complete spectrum
\begin{equation}
\frac{dE}{d E_\gamma} \propto 
\left \{
\begin{array}{lc} 
E_\gamma, & E_\rmscr{min} < E_\gamma <E_0  \\
E_\gamma^{-1/2}, & E_0
< E_\gamma <   E_\rmscr{break} \\
E_\gamma^{-2}, & E_\rmscr{break}  < E_\gamma < E_\rmscr{max}
\end{array}
\right .
\label{eq:24}
\end{equation}
The model has two free parameters: the total energy delivered by the
fast mode $E_\rmscr{total}$ and the strength of the magnetic field at
the inner edge of the pair-production region that determines the
location of both break points.

\begin{figure}
\includegraphics[width=80mm]{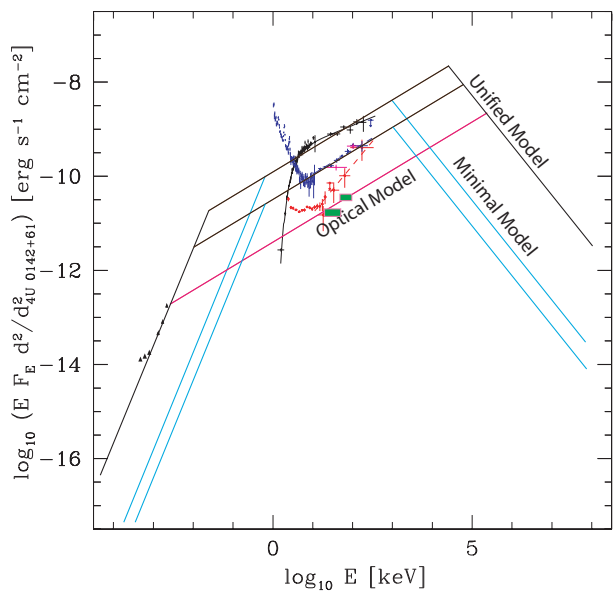}
\caption{The spectrum produced by fast-mode breakdown is superimposed
over the observed thermal and non-thermal emission from several AXPs
and SGRs for models that fit either the optical or INTEGRAL data
solely and one that fits both sets of data.  The unabsorbed optical
data are from Hulleman et al. \cite{2000Natur.408..689H} via
{\"O}zel \cite{2004astro.ph..4144O} for AXP 4U~0142+61.  The uppermost
black symbols are the hard X-ray band are from Molkov et
al. \cite{2004astro.ph.11696M} for SGR~1806-20.  Mereghetti et
al. \cite{2004astro.ph.11695M} obtained similar results for the SGR.
The middle sets of points in the hard X-ray data (blue is total flux
and red is pulsed flux) are from Kuiper et
al.\cite{2004ApJ...613.1173K} for AXP 1E~1841-045.  The green squares
plot the INTEGRAL data reported by den Hartog et al. \cite{2004ATel..293....1D} for
AXP~4U~0142+61. We normalised the den Hartog et al.\cite{2004ATel..293....1D} results
using the observations of the Crab by Jung \cite{1989ApJ...338..972J}.  We
scaled the emission from the three sources by assuming that they all
lie at the distance of AXP~4U~0142+61.  We used 3 kpc for
AXP~4U~0142+61 \cite{2000Natur.408..689H}, 7.5 kpc for AXP 1E~1841-045
\cite{1992AJ....104.2189S} and 15 kpc for SGR~1806-20
\cite{2004astro.ph.11696M}.}
\label{fig:allspec}
\end{figure}

Fig.~\ref{fig:allspec} compares the spectrum of non-thermal radiation
produce by fast-mode breakdown with the non-thermal radiation observed
from the AXPs 4U~0142+61 and 1E~1841-045 and the SGR 1806-20.   The
model accounts for the observed spectral slope both in the optical and
in the hard x-ray.  If we assume that all of the objects have similar
intrinsic optical emission to the nearby AXP 4U~0142+61, we predict
that the non-thermal emission should peak around 100~MeV as traced by
the curves denoted as the ``Optical Model'' for AXP 4U~0142+61 and as
the ``Unified Model'' for the other two objects.  On the other hand if 
1E~1841-045 and the SGR 1806-20 do not exhibit a similar optical
excess to 4U~0142+61, the non-thermal emission should extend to at
least 1~MeV as depicted by the ``Minimal Model'' curves.

EGRET determined upper limits for the gamma-ray flux from the
direction of these objects.  Tab.~\ref{tab:egret} gives these upper
limits and the predictions of the fast-mode breakdown model. The optical model and the unified
model predict approximately similar EGRET fluxes for 4U~0142+61.
The minimal model based solely on the INTEGRAL data exhibits a flux
above 100~MeV about two hundred times smaller than the unified model.
Because the optical model cannot explain the observed INTEGRAL data
for 1E~1841-045 and SGR~1806-20, it is omitted.  Similarly, the
minimal model cannot explain the optical data for 4U~0142+61.  We see
that for 1E~1841-045 and SGR~1806-20 the predictions for the minimal
model lie comfortably below the EGRET upper limits.  In the context of
the fast-mode breakdown model, this means that the optical emission
for 1E~1841-045 and SGR~1806-20 is inherently weaker than from
4U~0142+61.

On the other hand, 4U~0142+61 is difficult to explain in the context
of either model because of its large optical flux.  Perhaps 4U~0142+61
was more active during the epoch of the optical observations than
during the EGRET observations.  Some AXPs exhibit variable X-ray
emission such as AX J1845-0258 and 1E~1048.1-5937
\citep{2000ApJ...542L..49V,2004ApJ...608..427M} so this conclusion
might be natural.
\begin{table*}
\caption{Predicted flux above 100~MeV and observed EGRET upper limits \citep{Gren05}  in units of $10^{-8}$ photons s$^{-1}$cm$^{-2}$. The GLAST upper
  limits are nominally $(0.2~\rmmat{--}~ 0.4) \times 10^{-8}$  photons
  s$^{-1}$cm$^{-2}$ \citep{glast}.}
\label{tab:egret}
\begin{center}
\begin{tabular}{l|ccccc}
Object & EGRET            & EGRET       & Unified & Minimal & Optical \\
       & Exposure [weeks] & Upper Limit & Model   & Model   & Model 
 \\ \hline 
AXP~4U~0142+61  & 8.8     &   50        & 1500    & --- & 800 \\
AXP~1E~1841-045 & 6.8     &   70        &   70    & 0.4 & --- \\
SGR~1806-20     & 4.9     &   70        &  280    & 0.6 & --- \\
\end{tabular}
\end{center}
\end{table*}

\section{OUTLOOK}

Our understanding the magnetars has increased dramatically over the
past few years but so have the unknowns.  The recently discovered hard
x-ray emission
\cite{2004astro.ph.11696M,2004astro.ph.11695M,2004ApJ...613.1173K,2004ATel..293....1D}present
a theoretical challenge to understand but may also hold that key to
understanding the evolution of the mangetic field on magnetars that
drives not only the non-thermal and thermal quiescent emission but the
bursts as well.  The recent massive burst from SGR~1806-20
\cite{2005astro.ph..3030P,2005astro.ph..2577M,2005astro.ph..2541M}
shows that the size of soft-gamma repeater bursts varies over two
additional orders of magnitude.  The gamma-ray energy from the
December 27 event ($2\times 10^{46}$~erg) is a whopping two-percent of
the total electromagnetic energy from a supernova, and the SGR is
expected to burst like this many times over its lifetime.  The
observation of SGR-like bursts from AXPs has further unified these two
classes of objects and the association of glitches with the
bursts\cite{2003ApJ...588L..93K} from AXP 1E~2259+586 gives a
tantalizing hint at the underlying mechanism for these bursts.

We still do not have a adequate theoretical description of the thermal
spectra from AXPs and SGRs.  Where do the expected line features go?
Although features have been seen in 1RXS J170849-400910
\cite{2003ApJ...586L..65R}, the other AXPs and isolated neutron stars
such as RX J185635-3754 and RX J0720.4-3125 exhibit no features.  Is
this an indication that magnetars have condensed surfaces
\cite{2004astro.ph..6001V} or do they have helium, carbon or iron
atmospheres that have not yet been studied in the magnetar regime?

Understanding magnetars draws upon a wide range of physical processes
from quantumelectrodynamics to plasma physics, from condensed matter
physics to general relativity.  The physics is messy, but that makes
it fun!
 
\bigskip 
\begin{acknowledgments}
The National Science and Engineering
Research Council of Canada supported this work.  J. S. Heyl is a
Canada Research Chair.
\end{acknowledgments}

\bibliographystyle{prsty}
\bibliography{ns,qed,mine,physics}

\end{document}